\documentclass[a4paper,11pt]{article}
\pdfoutput=1 

\usepackage{jheppub} 

\usepackage[T1]{fontenc} 

\usepackage{amsfonts}
\usepackage{graphics}
\usepackage{epsfig}
\usepackage{url}
\usepackage{multirow}
\usepackage{feynmp}
\newcommand {\be}{\begin{equation}}
\newcommand {\ee}{\end{equation}}
\newcommand {\ba}{\begin{eqnarray}}
\newcommand {\ea}{\end{eqnarray}}

\newcommand {\tanb}{$\tan\beta~$}
\newcommand {\ra}{\rightarrow}
\newcommand {\mh}{$m(H^{\pm})$}

\title{\boldmath Observability of Heavy Charged Higgs through $s$-channel Single Top Events at LHC }

\author[]{M. Hashemi}

\affiliation[]{Physics Department and Biruni Observatory,\\College of Sciences, Shiraz University, Shiraz 71454, Iran}

\emailAdd{hashemi$\_$mj$@$shirazu.ac.ir}

\abstract{
The charged Higgs bosons can be produced as a resonance in $s$-channel single top events. The light charged Higgs in such events preferably decays to a pair of $\tau\nu$ thus making it difficult to distinguish from the large single $W$ events producing the same final state. However, the heavy charged Higgs decay to a pair of $t\bar{b}$ can be extracted from the SM background events. The final state under consideration in this paper contains the top quark decay to $W^{+}b$ followed by $W^+$ decay to electrons or muons. It is shown that this signal can be observed at LHC at a large area of MSSM phase space $(m(H^{\pm}),\tan\beta)$. Finally $5\sigma$ and 95$\%$ CL exclusion contours are presented at different integrated luminosities of LHC assuming a nominal center of mass energy of $\sqrt{s}=14$ TeV.}

\begin{document} 
\maketitle
\flushbottom

\section{Introduction}
The charged Higgs bosons of the Minimal Supersymmetric Standard Model (MSSM) provide a unique signature of a theory beyond the Standard Model (SM) due to their electric charge which makes them different from the neutral SM Higgs boson in terms of their production, interaction and decay properties. Therefore there have been extensive searches for this particle during the last years at Tevatron (Fermilab) and LEP and LHC (CERN). \\
The MSSM parameter space is usually quoted in terms of two parameters: \mh and \tanb which is the ratio of vacuum expectation values of the two Higgs fields used to make the two Higgs doublets. \\
The LEP II direct searches set a lower limit on the charged Higgs mass as $m(H^{\pm})>89~ \textnormal{GeV}$ \cite{lepexclusion1}. This limit is in contrast to the stronger limit, $m(H^{\pm})>125~ \textnormal{GeV}$, obtained from indirect searches at LEP \cite{lepexclusion2}. The Tevatron searches by the D0 \cite{d01,d02,d03,d04} and CDF Collaborations \cite{cdf1,cdf2,cdf3,cdf4} exclude high \tanb values in the light charged Higgs area. The recent limits from LHC experiments (analyses of $\textnormal{H/A}\rightarrow \tau \tau$) exclude high \tanb for a charged Higgs boson in the mass range $200<m_{H^{\pm}}<400$ GeV \cite{cmsindirect,atlasindirect}. However, these limits are not yet confirmed by direct searches for charged Higgs from ATLAS \cite{atlasdirect} and CMS \cite{cmsdirect,cmsdirect2} as they cover only light charged Higgs up to $m_{H^{\pm}}=160$ GeV. From these results, a high \tanb as high as 50 is still allowed. A recent result from ATLAS \cite{atlasdirect2} excludes high \tanb for $m_{H^{\pm}}>200$ GeV, however, the assumption BR($H^{\pm}\rightarrow \tau\nu$)=1, used in this analysis, is not the case for heavy charged Higgs and makes exclusions overestimated. There are stronger limits from flavor physics. An indirect search using CLEO data requires \mh$>295 ~\textnormal{GeV}$ at 95 $\%$ C.L. in 2HDM Type II with \tanb$>2$ \cite{B1}, such a constraint is not considered as it belongs to general 2HDMs withought supersymmetric constraints.\\
The purely leptonic decay $B_s \rightarrow \mu^+\mu^-$ is one of the processes which are sensitive to supersymmetric contributions. The present measurement of BR($B_s \rightarrow \mu^+\mu^-$) at LHCb and its implication for constrained and unconstrained MSSM, has been studied in detail in \cite{bs} where it is shown that including results from LHC searches for SUSY and Higgs, the achieved accuracy on BR($B_s \rightarrow \mu^+\mu^-$) leaves a large fraction of SUSY parameter space, unconstrained.  The two types of MSSM, i.e., CMSSM and pMSSM have been proved to exclude only about 11$\%$ and 3$\%$ of the valid points in parameter space which are not yet excluded by LHC searches for SUSY. The remaining large part of the parameter space is thus allowed by current  measurement of BR($B_s \rightarrow \mu^+\mu^-$).\\
As a summary, although, \tanb values relevant to this analysis are not favored by the LHC searches for neutral Higgs bosons of MSSM, results from direct searches for the charged Higgs are used as the baseline for this work. The output of the analysis would most likely be useful for a direct exclusion and a confirmation of LHC results from searches for neutral Higgs bosons, although a discovery potential is also discussed in the last section.  
\section{Light and Heavy Charged Higgs Bosons Produced in Single Top Events}
The single top events are produced in $s$, $t$ and $tW$ channels. A comprehensive review of the top quark physics including a description of single top events can be found in \cite{tp}. The D0 \cite{d0st} and CDF \cite{cdfst} were the first collaborations which announced observation of single top events . These reports were followed by those of the ATLAS and CMS collaborations on the observation and cross section measurement of these events \cite{atlasst,cmsst}. \\
In addition to analyses of single top as an SM electroweak process which provides an opportunity to estimate $V_{tb}$ element of the CKM matrix, these events have been studied as sources of the charged Higgs bosons. In \cite{oldreview,st1,st2}, single top events were used as a source of light charged Higgs decaying to $\tau\nu$ in the leptonic final state ($\tau$ leptonic decay). The hadronic final state has been studied in a previous work in \cite{myjhep}. The above two types of analyses deal with light charged Higgs in decays of top quarks produced in a $t$-channel single top. Therefore the domain of the analysis is limited to the light charged Higgs, i.e., \mh$<m_{\textnormal{top}}$.\\
On the contrary, the heavy charged Higgs can be analyzed through the $s$-channel single top production assuming that it contributes to the $s$-channel propagator. In \cite{schcms,schatlas}, the $s$-channel charged Higgs production is analyzed in the $\tau\nu$ and $t\bar{b}$ final states respectively. Of course the first process is not a single top process anymore, however it is a source of light charged Higgs in the $\tau\nu$ final state. In \cite{st3}, a similar analysis to what has been reported in \cite{schatlas} has been performed by Tevatron. The above analyses rely on TopReX generator \cite{toprex} for the signal generation and cross section calculation. This package overestimates the signal cross section due to the use of leading order calculations for decay rates. On the other hand, the single W decay to $\tau\nu$ is underestimated in \cite{schcms}. In \cite{schatlas}, the charged Higgs invariant mass is not used in the chain of selection cuts. In this paper, a mass window cut is applied on the charged Higgs candidate invariant mass and the signal to background ratio is enhanced. The report presented in \cite{st3} is based on an analysis relevant to Tevatron center of mass energy and results are superseded by those obtained in this paper.\\
There is also a discussion of the single charged Higgs production in \cite{0708}, where, it is shown that squark mixing can significantly change the production rate. They discuss the observability of a single charged Higgs or a charged Higgs in association with a hard jet in the light of flavor physics and show that extended flavor structures can have a sizable effect on the production rate. These effects are not included in this paper, however, the tree level results quoted in \cite{0708} are reproduced with our calculation and a reasonable agreement is obtained.\\
As an example, the cross section for $H^+$ production with a mass $m_{H^{\pm}}=188$ GeV at \tanb = 7, is quoted as 41.2 $fb$ in \cite{0708}. A calculation using CompHEP gives $\sigma(pp \rightarrow t\bar{b})=13.5~fb$. The HDECAY, (with 2-loop level decay rates), gives BR$(H^{\pm} \rightarrow t\bar{b})=0.35$, which leads to $\sigma(H^+)=38.6~fb$. This small cross section increases with increasing \tanb. Assuming a $\tan^2 \beta$ dependence of the cross section, such cross section would be $O(1~pb)$ at \tanb =50 or higher.\\
In the following sections, signal and background events are introduced and their cross sections are presented. An event selection and analysis is described in detail with the aim of charged Higgs invariant mass reconstruction with different mass hypotheses. Finally an estimation of accessible regions of MSSM parameter space (\mh,\tanb) for a $5\sigma$ discovery or exclusion at 95 $\%$ CL is provided. The analysis is based on MSSM, $m_{h}$-max scenario with the following parameters: $M_{2}=200$ GeV, $M_{\tilde{g}}=800$ GeV, $\mu=200$ GeV and $M_{SUSY}=1$ TeV.

\section{Signal and Background Processes and Their Cross Sections}
\subsection{Signal}
\begin{figure}
 \centering
 \includegraphics[width=0.6\textwidth]{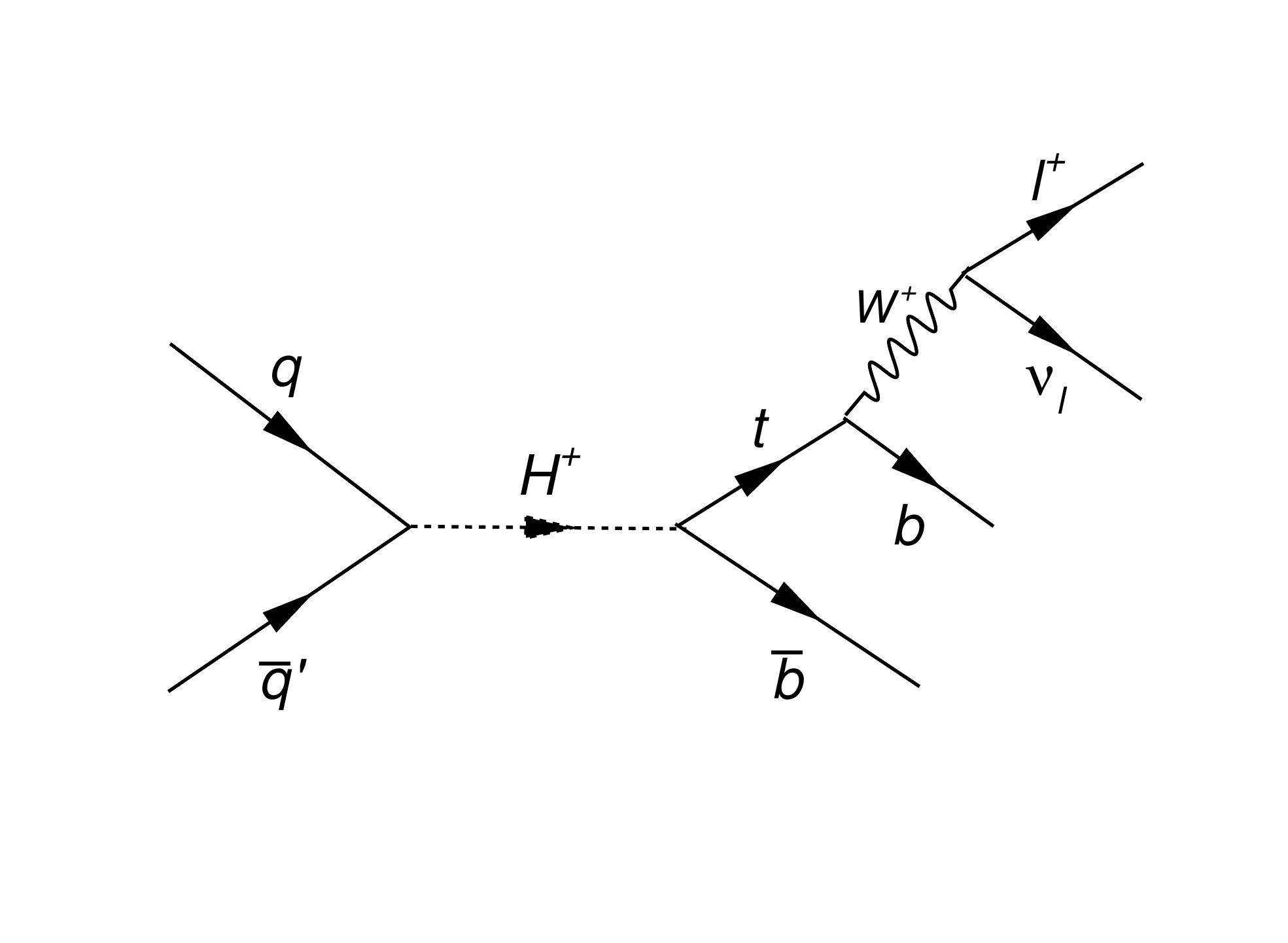}
 \caption{The full signal production chain: the $s-$channel single top diagram as the signal process followed by the top quark decay to $W$ boson which in turn decays leptonically.\label{diagram}}
 \end{figure}

\begin{figure}
 \centering
 \includegraphics[width=0.7\textwidth]{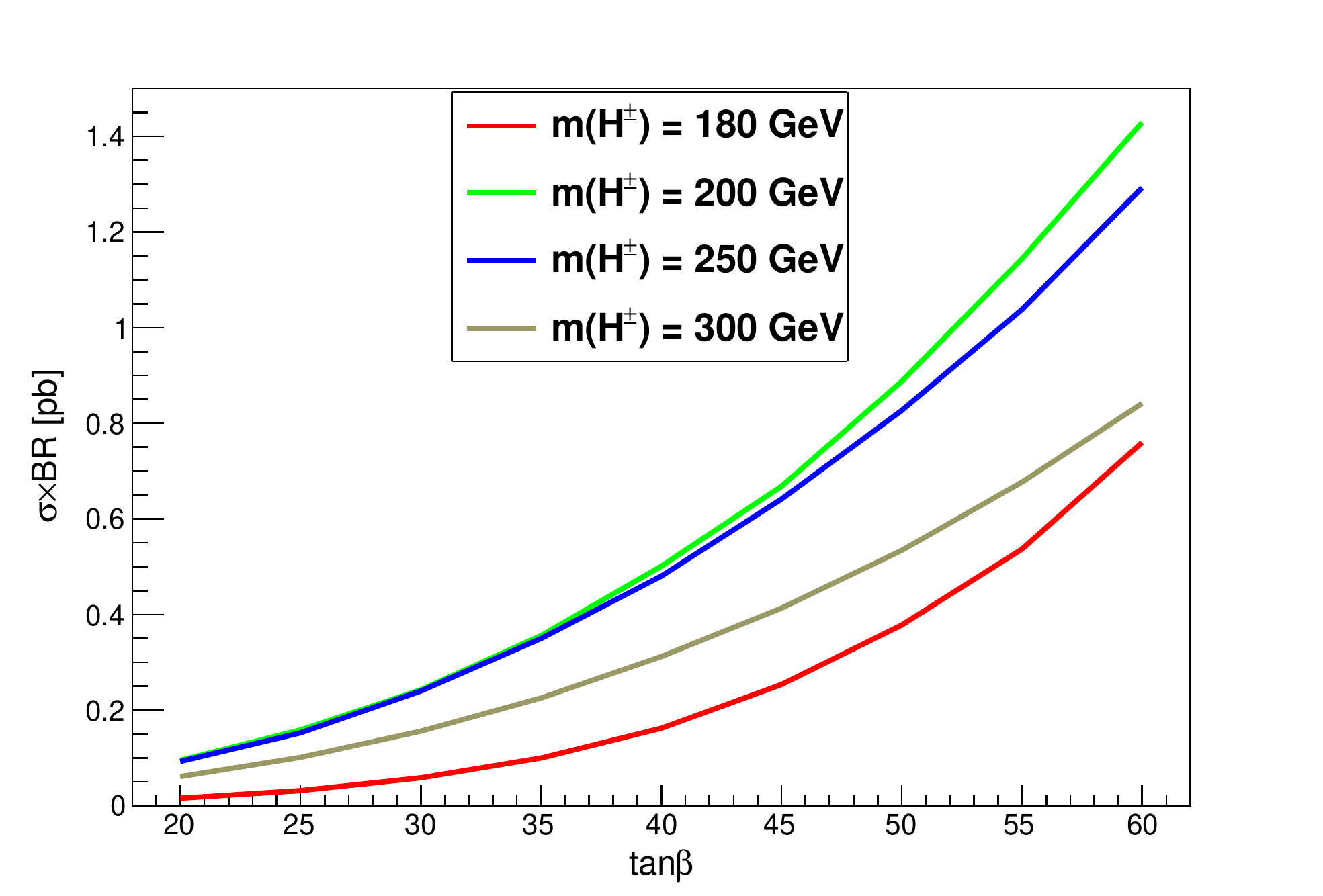}
 \caption{The signal cross section times branching ratio as a function of the charged Higgs mass and \tanb. The used branching ratios are $BR(t\ra Wb)\simeq 1$ and $BR(W\ra \ell\nu)\simeq0.2$. \label{signalxsec}}
 \end{figure}

\begin{figure}
 \centering
 \includegraphics[width=0.7\textwidth]{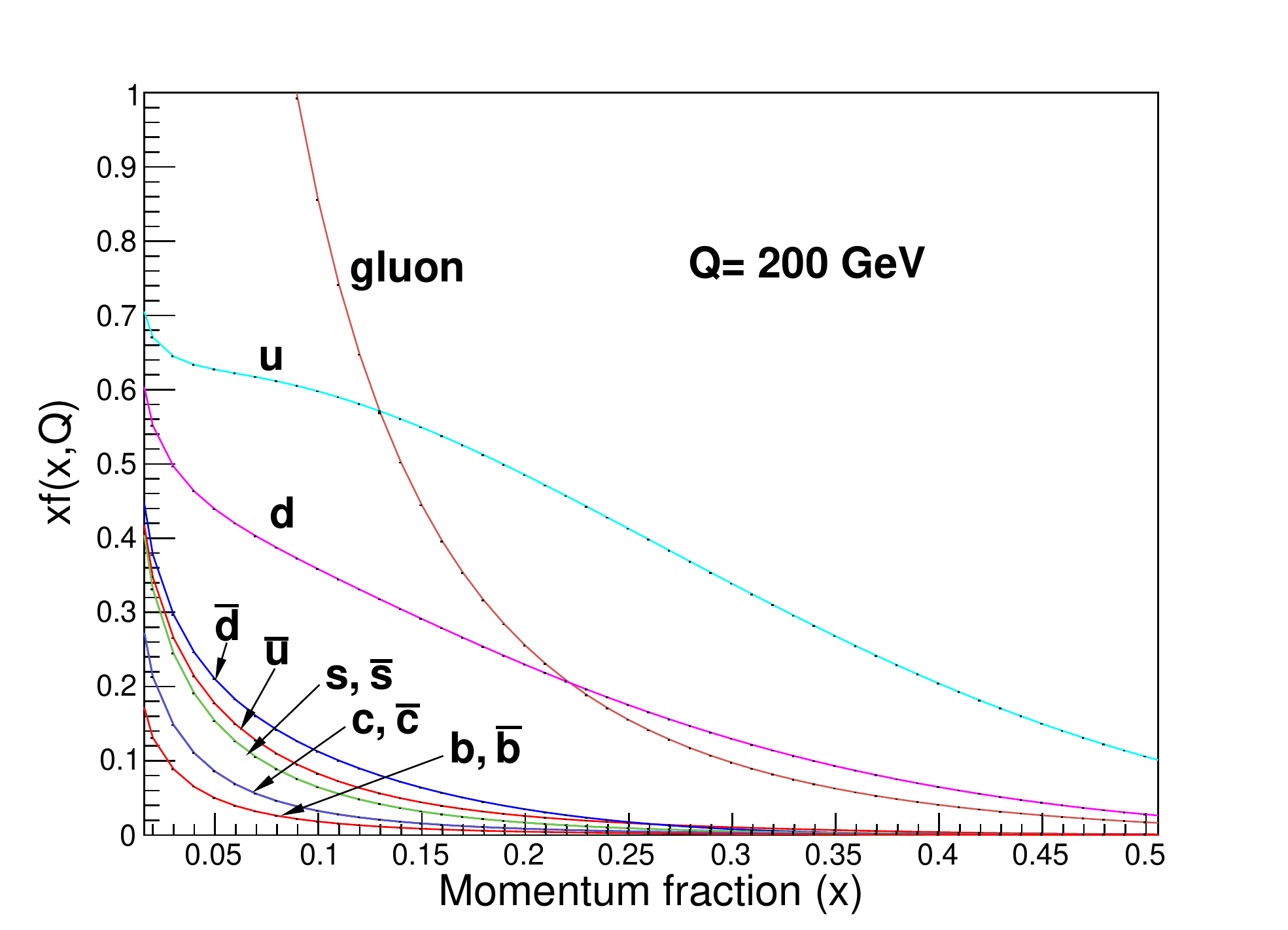}
 \caption{The parton distribution functions at $Q=200$ GeV, using CTEQ 6.6. \label{pdfs}}
 \end{figure}

\begin{figure}
 \centering
 \includegraphics[width=0.7\textwidth]{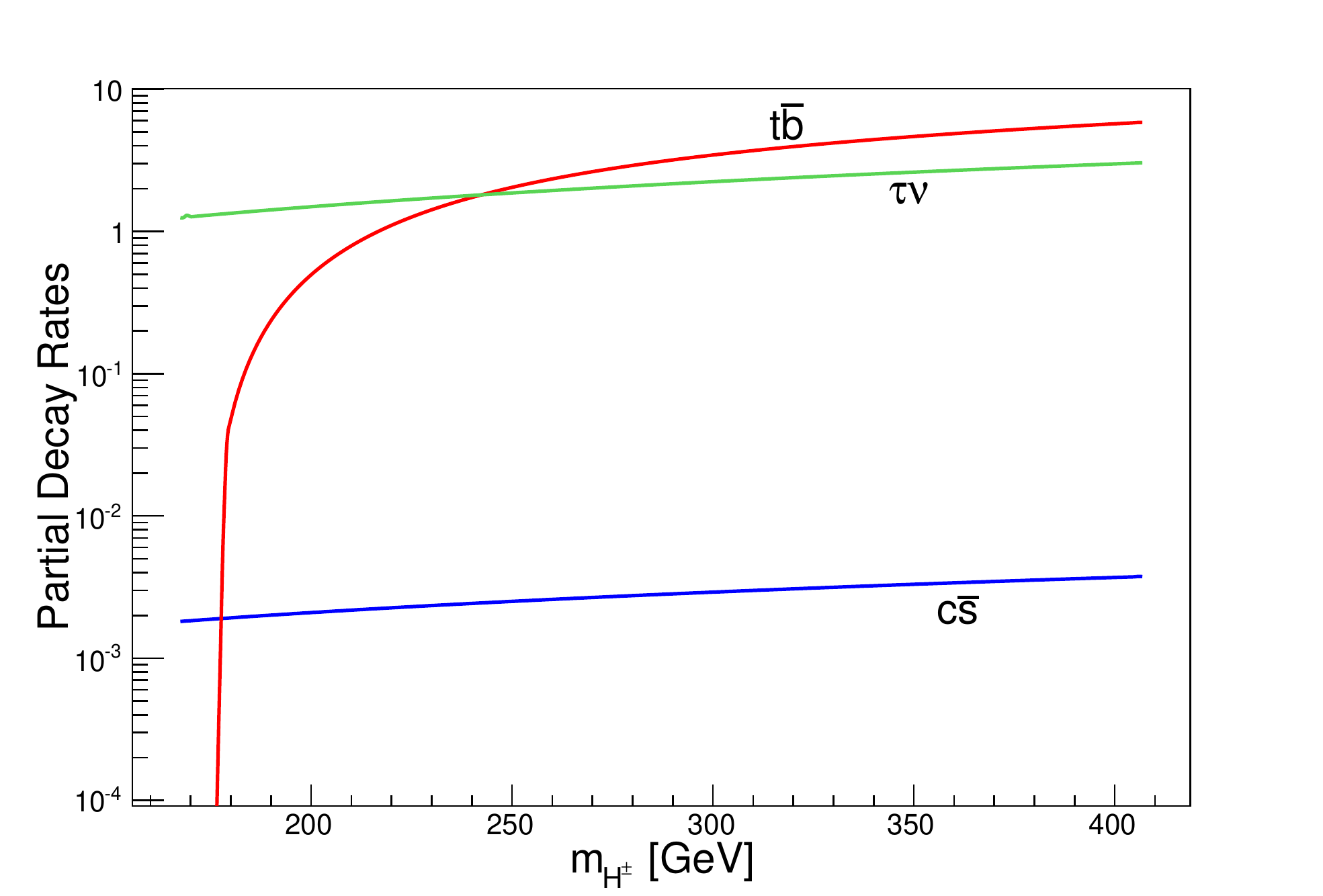}
 \caption{The charged Higgs partial decay rates at \tanb = 60. \label{widths}}
 \end{figure}

The signal is the $s$-channel single top with a charged Higgs in the intermediate phase (propagator) decaying to a $t\bar{b}$ quark pair as in Fig. \ref{diagram}. The $W$ boson in the top quark decay undergoes a leptonic decay to an electron or a muon.
The cross sections of signal ($\ell^+\nu b\bar{b} + \ell^-\bar{\nu} b\bar{b}$ final state) with different charged Higgs mass and \tanb values are shown in Fig. \ref{signalxsec}. Values shown in Fig.\ref{signalxsec} are not yet excluded by LHC experiments \cite{atlasdirect,cmsdirect}. The used package is CompHEP 4.5.1 \cite{comphep,comphep2} with the following input values for the quark masses: $m_c=1.3$ GeV, $m_s=0.1$ GeV, $m_t=173$ GeV and  $m_b=4.8$ GeV. The charged Higgs total decay rate is taken from HDECAY 4.1 \cite{hdecay} and CTEQ 6.6, provided by LHAPDF 5.8.3 \cite{lhapdf}, is used as the parton distribution function. This PDF is used throughout the analysis.
An independent calculation is also performed using CTEQ 6.6 and the general cross section formula of the $s$-channel annihilation in quark-antiquark scattering as in Eq. \ref{sigmahat} (\cite{sxsec}).
\be
\hat{\sigma}=\frac{16\pi m^2_{H^{\pm}}}{\hat{s}}\frac{\Gamma(H^{\pm}\rightarrow c\bar{s})\Gamma(H^{\pm}\rightarrow t\bar{b})}{(\hat{s}-m^2_{H^{\pm}})^2+m^2_{H^{\pm}}\Gamma^2_{\textnormal{total}}}
\label{sigmahat}
\ee
Here $\hat{s}=x_i x_j s$, where $s$ is the square of the center of mass energy ($\sqrt{s}=14$ TeV) and $x_i$ and $x_j$ are proton momentum fractions carried by the two partons which are involved in the interaction. Therefore the partonic cross section, $\hat{\sigma}$, depends on $x_i$ and $x_j$. It should be noted that in writing Eq. \ref{sigmahat}, no spin factor has been assumed due to the fact that the charged Higgs is spinless. This feature imposes a constraint on the incoming partons spins. They have to be in the spin singlet configuration. The proton-proton cross section is then obtained by inserting $\hat{\sigma}$ (Eq. \ref{sigmahat}) in Eq. \ref{sigmatot} which takes into account the parton distribution functions.
\be
\sigma=\sum_{i,j}\int dx_i \int dx_j ~f(x_i,Q,i) f(x_j,Q,j) ~\hat{\sigma}
\label{sigmatot}
\ee   
In Eq. \ref{sigmatot}, $i$ and $j$ denote the parton flavor. The incoming quarks effectively make only one of the pairs $c\bar{s}$, $s\bar{c}$, $\bar{s}c$ and $\bar{c}s$. Since $c$, $\bar{c}$, $s$ and $\bar{s}$ are seq quarks, their momentum distributions are effectively equivalent. This effect is observed in Fig. \ref{pdfs}. Therefore Eq. \ref{sigmatot} can be evaluated as four times the cross section of $c$-$\bar{s}$ interaction. 
The partial decay rates in the numerator of Eq. \ref{sigmahat} and the total decay rate of the charged Higgs are obtained using HDECAY as shown in Fig. \ref{widths} for a typical value of \tanb = 60. As is seen from Fig. \ref{widths}, the charged Higgs decay rate to $c\bar{s}$ is very small leading to a small partonic cross section $\hat{\sigma}$. However, when $\hat{\sigma}$ is multiplied by parton distribution functions and integrated over all possible momentum fraction values and incoming partons, the final result is sizable. As an example, at $m_{H^{\pm}}=200$ GeV and \tanb = 60, using $\Gamma(H^{\pm}\rightarrow c\bar{s})=0.002$ GeV, $\Gamma(H^{\pm}\rightarrow t\bar{b})=0.5$ GeV and $\Gamma_{\textnormal{total}}=2.02$ GeV, the signal cross section ($\sigma \times$ BR$(W\rightarrow \ell \nu)$) is obtained as 1.4 $pb$ which is in a reasonable agreement with what is obtained using CompHEP (shown in Fig. \ref{signalxsec}).
\subsection{Background}
The main background processes are $W^{\pm}jj$, $W^{\pm}b\bar{b}$ and $W^{\pm}c\bar{c}$ generated with ALPGEN \cite{alpgen} and $t\bar{t}$, $s$-channel and $t$-channel single top events generated with PYTHIA 8.153 \cite{pythia}. Table \ref{xsec} shows the cross section of background processes as well as the tools used for their generation and cross section calculation. More information about the MCFM can be found in \cite{mcfm}.
The jet reconstruction has been performed with FASTJET 2.4.1 \cite{fastjet} using anti-kt algorithm \cite{antikt} and $E_{T}$ recombination scheme with a cone size of $\Delta R=0.5$ where $\Delta R=\sqrt{(\Delta\eta)^2+(\Delta\phi)^2}$. Here $\eta=-\ln\tan\theta/2$ and $\theta (\phi)$ is the polar (azimuthal) angle with the beam pipe defined as the $z$-axis.    
\begin{table}[h]
\begin{center}
\begin{tabular}{|c|c|c|c|c|}
\hline
\multirow{2}{*}{Process} & Generation & Cross section &  Kinematic & \multirow{2}{*}{$\sigma \times $BR [pb]}  \\
 & tool & tool & preselection &  \\
\hline
$t\bar{t}$ & PYTHIA & MCFM & - & 227 \\
\hline
SM $s$-channel & \multirow{2}{*}{PYTHIA} & \multirow{2}{*}{MCFM} & \multirow{2}{*}{-} & \multirow{2}{*}{2.1} \\
single top & & & & \\
\hline
SM $t$-channel & \multirow{2}{*}{PYTHIA} & \multirow{2}{*}{MCFM} & \multirow{2}{*}{-} & \multirow{2}{*}{50} \\
single top & & & & \\
\hline
\multirow{2}{*}{$Wjj$} & \multirow{2}{*}{ALPGEN} & \multirow{2}{*}{ALPGEN} &  $E_T^{\textnormal{jet}}>20$ GeV & 3340 \\
& & & $\eta^{\textnormal{jet}}<5$ & \\
\hline
\multirow{2}{*}{$Wb\bar{b}$} & \multirow{2}{*}{ALPGEN} & \multirow{2}{*}{ALPGEN} &  $E_T^{\textnormal{jet}}>20$ GeV & 7 \\
& & & $\eta^{\textnormal{jet}}<5$ & \\
\hline
\multirow{2}{*}{$Wc\bar{c}$} & \multirow{2}{*}{ALPGEN} & \multirow{2}{*}{ALPGEN} &  $E_T^{\textnormal{jet}}>20$ GeV & 6.6 \\
& & & $\eta^{\textnormal{jet}}<5$ & \\
\hline
\end{tabular}
\end{center}
\caption{Cross sections of background processes. Values shown are the total cross sections times branching ratio of decays which produce the final state related to the analysis. Therefore the semi-leptonic $t\bar{t}$ is analyzed and for single top, $Wjj$, $Wb\bar{b}$ and $Wc\bar{c}$, the $W$ leptonic decay is considered. The lepton under consideration is only $e$ and $\mu$. The $\tau$ lepton is not considered due to its soft leptonic decay.  \label{xsec}}
\end{table}

\section{Event Selection and Analysis}
The signal event shown in Fig. \ref{diagram} indicates that the final state to analyze consists of two $b$-jets and a lepton ($e$ or $\mu$) and some missing $E_T$ (denoted as $E^{\textnormal{miss}}_T$). The existence of two $b$-jets in the event is expected to dramatically suppress the large $Wjj$ sample. However, as will be seen, the $t\bar{t}$ events are the main background. In what follows a detailed description of the analysis and event selection strategy is presented.
\subsection{Selection Strategy}
The event selection starts with requiring one lepton and two jets passing the $b$-tagging algorithm. The $b$-tagging requirement is a simple matching based algorithm based on selecting jets which are close to $b$ or $c$ quarks, i.e., $\Delta R_{j,q}<0.4$. Here, $q=b$ or $c$. The selection efficiency is taken to be 60$\%$ for $b$-jets and 10$\%$ for $c$-jets.\\
In the next step, the $E^{\textnormal{miss}}_T$ and lepton components are used for $W$ four-momentum and invariant mass reconstruction. The $z$-component of the neutrino momentum ($p^{\nu}_z$) is constructed so as to give a right value for the $W$ mass in the $\ell\nu$ combination, i.e., $m_{\ell\nu}=80$ GeV. If such a solution is not found, $p^{\nu}_z$ is set to zero. In rare cases this situation occurs, resulting in a $W$ candidate mass different from the nominal value. Therefore a mass window on the $W$ candidate invariant mass is applied. Using the $W$ boson four-momentum, the right $b$-jet from the top quark decay is found by calculating the top quark invariant mass through the $\ell\nu b$ combination and finding the $b$-jet which gives the closest top quark mass to the nominal value, i.e., $m_t=173$ GeV. A mass window is further applied on the top quark candidate invariant mass distribution. A cut on $E^{\textnormal{miss}}_T$ is finally applied to further increase the signal to background ratio.\\
\subsection{Selection cuts}
\begin{figure}
 \centering
 \includegraphics[width=0.7\textwidth]{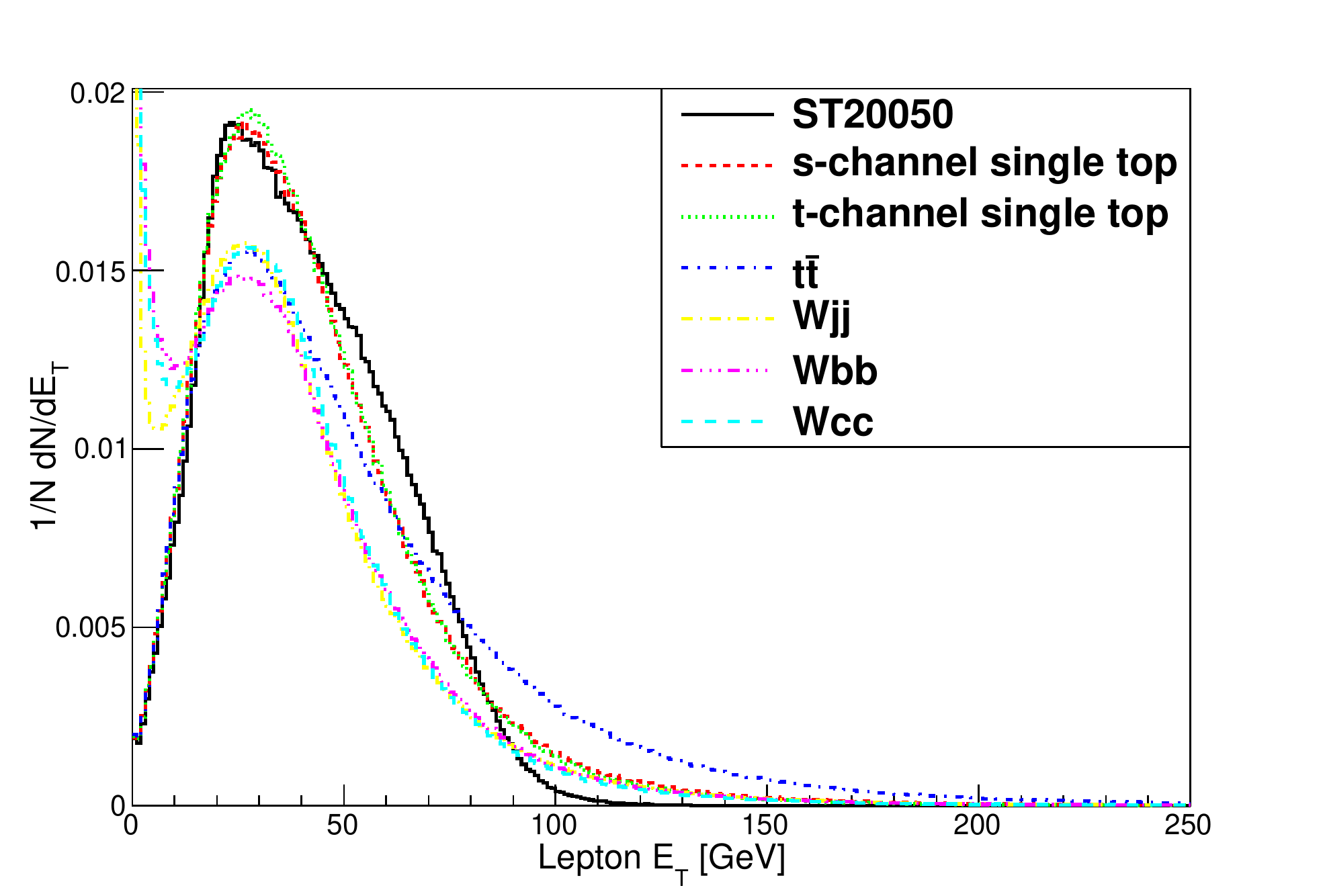}
 \caption{The lepton transverse momentum distribution in signal and background events. \label{let}}
 \end{figure}

\begin{figure}
 \centering
 \includegraphics[width=0.7\textwidth]{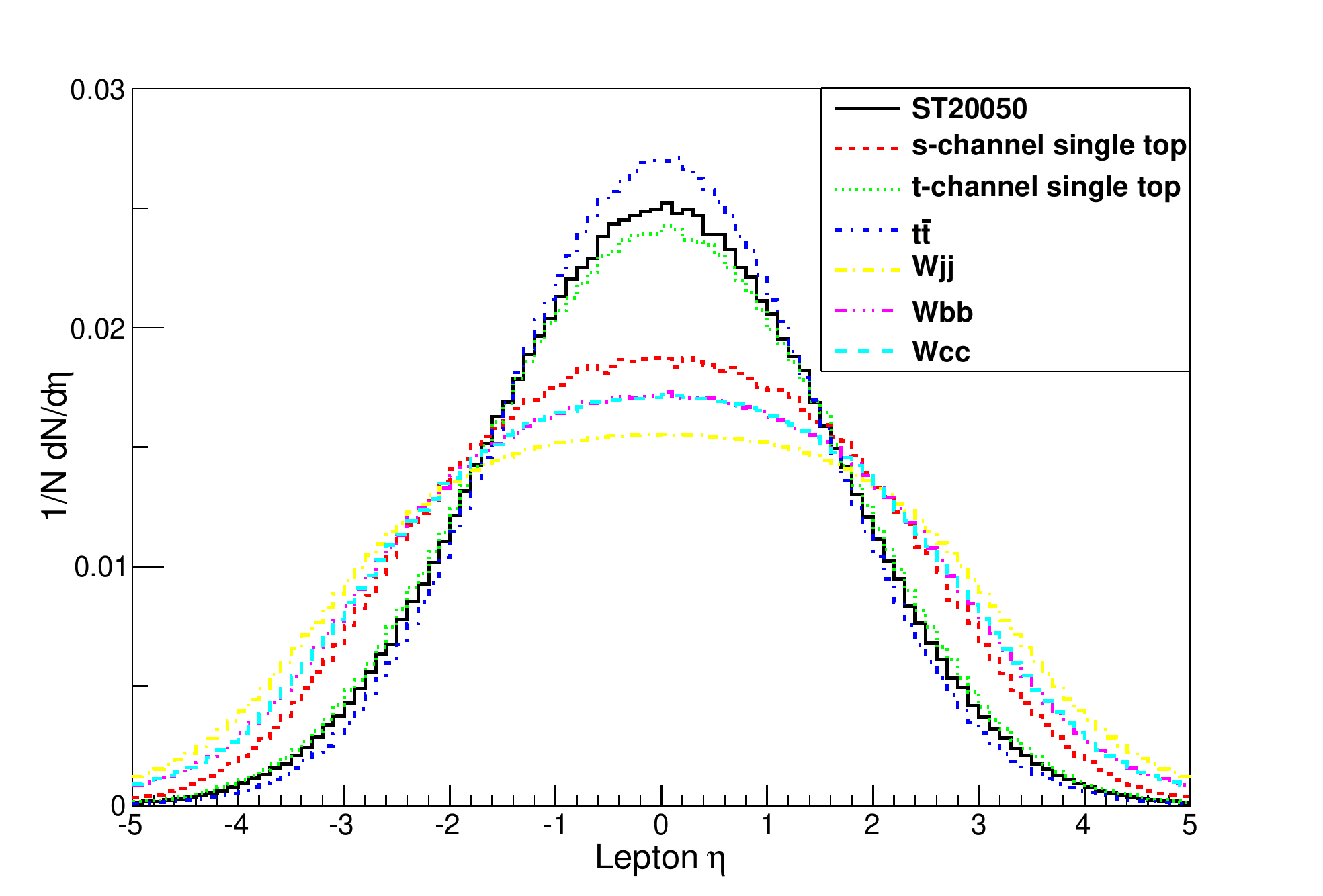}
 \caption{The lepton pseudorapidity distribution in signal and background events.\label{leta}}
 \end{figure}

\begin{figure}
 \centering
 \includegraphics[width=0.7\textwidth]{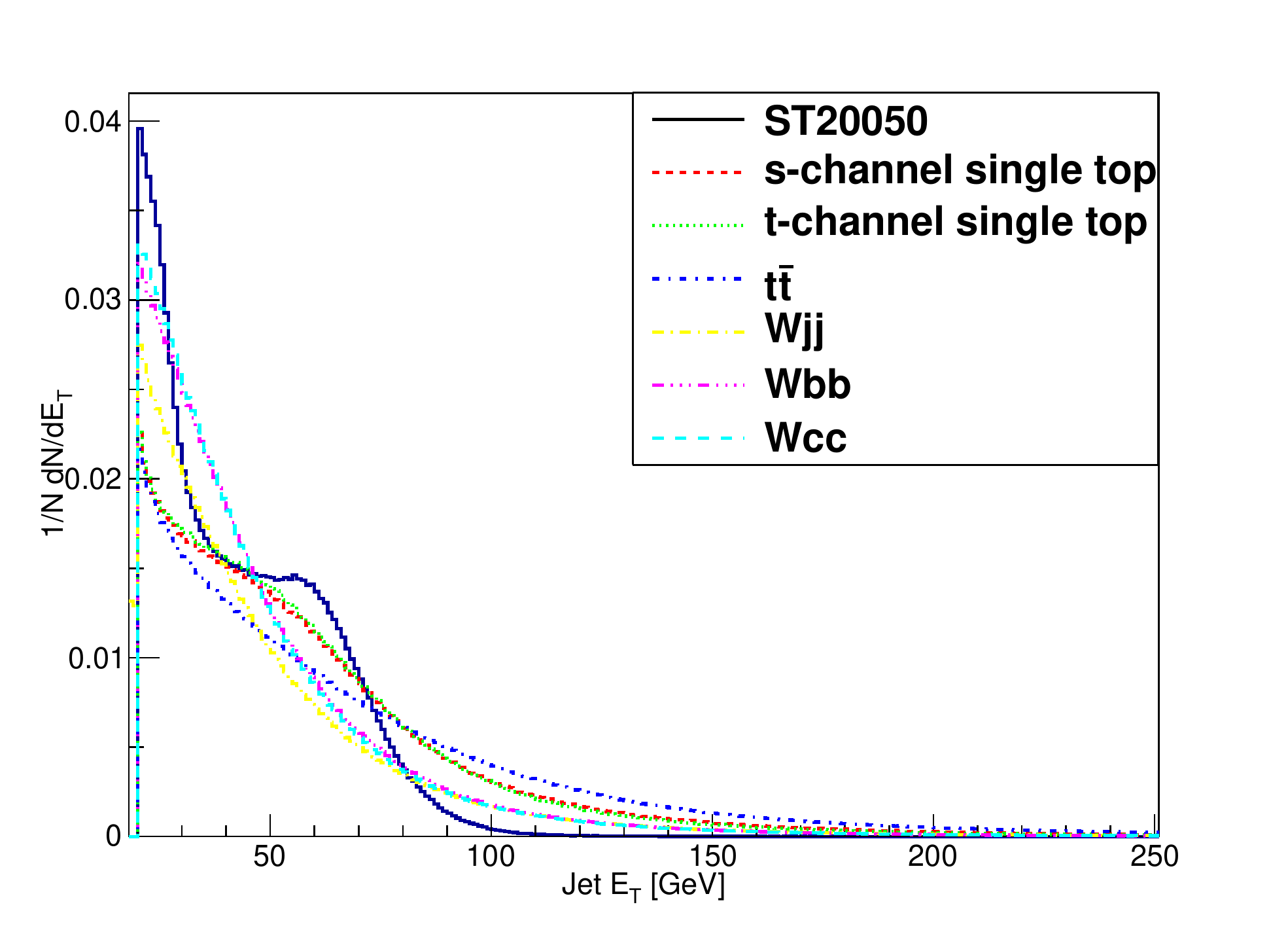}
 \caption{The jet transverse energy distribution in signal and background events. \label{jet}}
 \end{figure}

\begin{figure}
 \centering
 \includegraphics[width=0.7\textwidth]{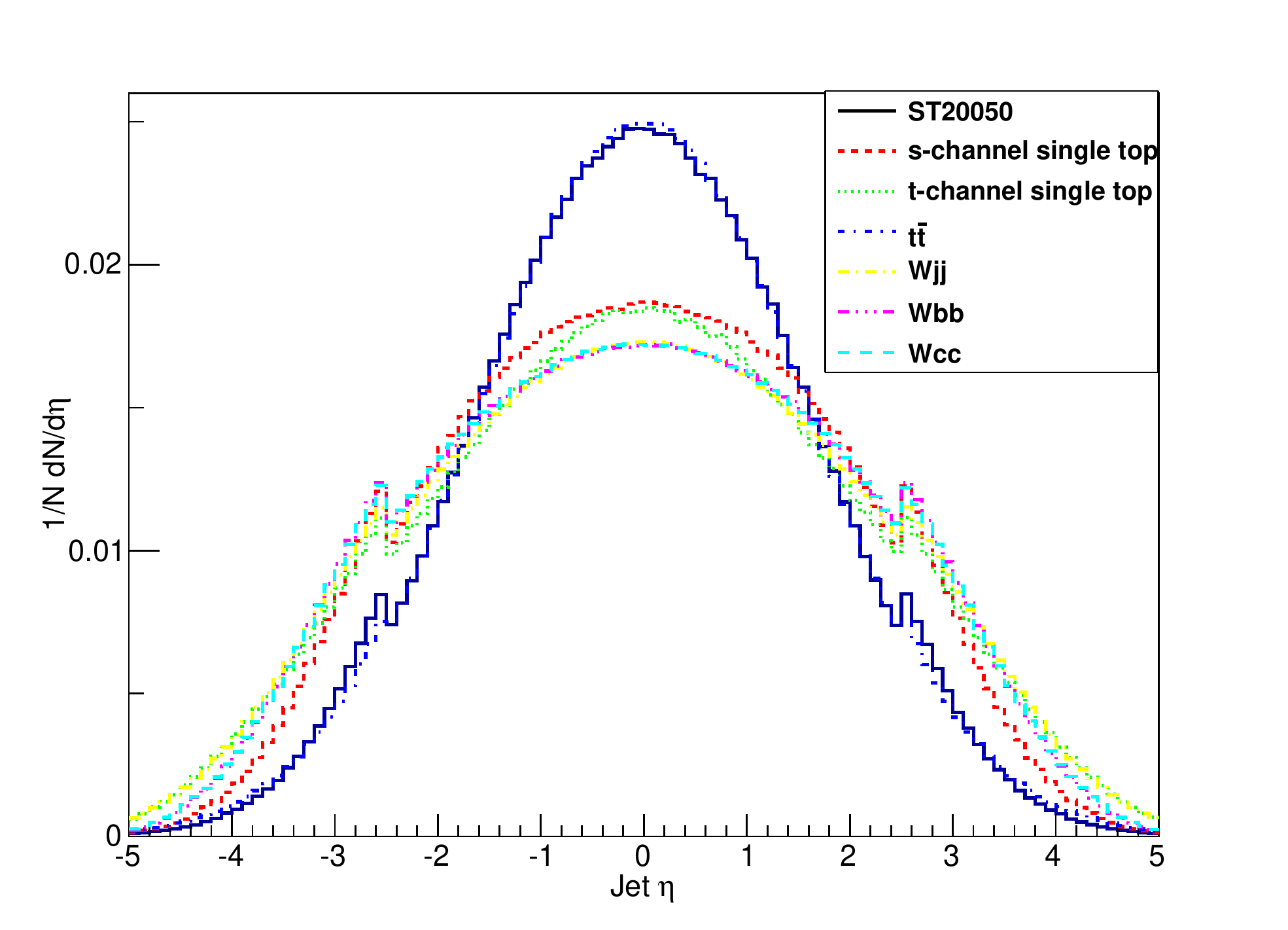}
 \caption{The jet pseudorapidity distribution in signal and background events. \label{jeta}}
 \end{figure}

\begin{figure}
 \centering
 \includegraphics[width=0.7\textwidth]{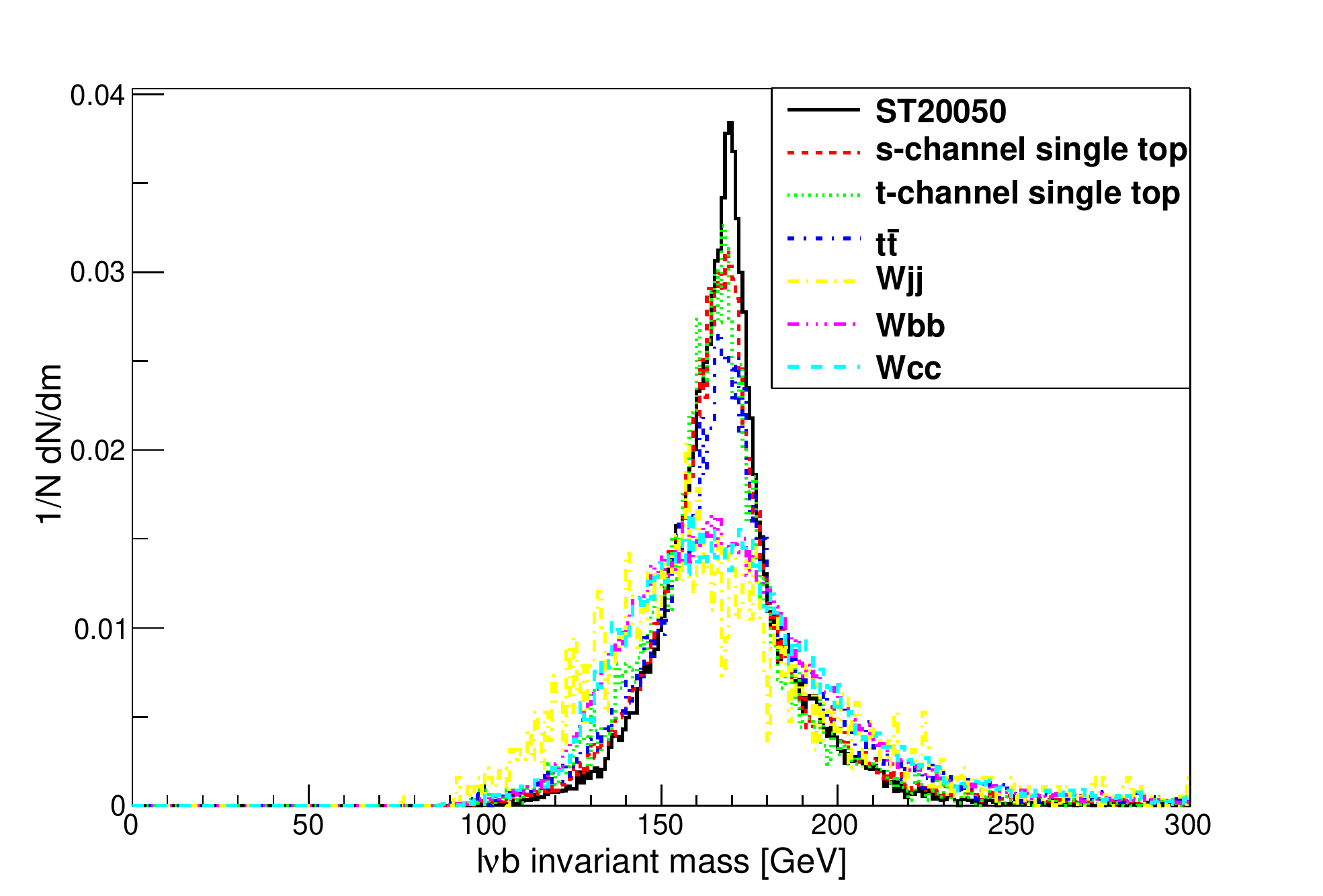}
 \caption{The top quark candidate invariant mass calculated from the $\ell\nu b$ four-momentum combination. \label{topmass}}
 \end{figure}

\begin{figure}
 \centering
 \includegraphics[width=0.7\textwidth]{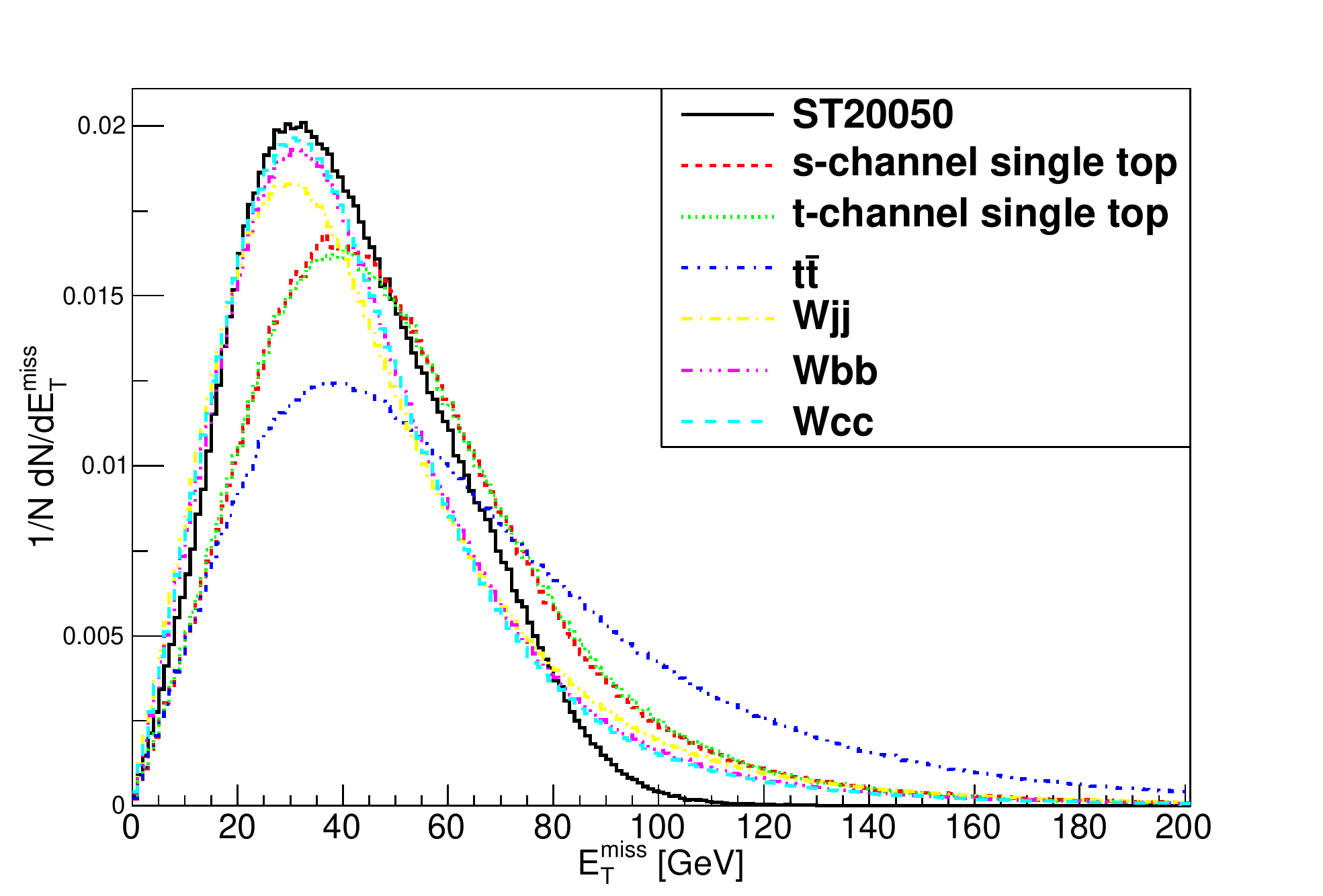}
 \caption{The $E^{\textnormal{miss}}_T$ distribution in signal and background events. \label{met}}
 \end{figure}

In order to select signal events and suppress the background samples, a reasonable understanding of the event kinematics is required. The selection cuts are applied according to kinematic and topological differences between the signal and background samples. To achieve this goal, the main kinematic distributions are plotted for signal and background samples. A comparison of such distributions would manifest the way the signal events can be distinguished from the background events. In what follows, the kinematic distributions of signal events with $m_{H^{\pm}}=200$ GeV and \tanb= 50 (abbreviated as ``ST20050``) are shown for illustration. The \tanb value is of course irrelevant for such distributions. It only contributes to the signal cross section. In Figs. \ref{let} and \ref{leta}, the lepton transverse momentum and pseudorapidity distribution is plotted for signal and background events. Similar to leptons, Figs. \ref{jet} and \ref{jeta} show jets transverse energy and pseudorapidity distributions. The top quark candidate invariant mass distribution is shown in Fig. \ref{topmass} and finally the missing $E_T$ distribution is plotted in Fig. \ref{met}. According to Figs. \ref{let},\ref{leta},\ref{jet},\ref{jeta},\ref{topmass} and \ref{met}, the following selection cuts listed in Tab. \ref{cuts} are applied.

\begin{figure}
 \centering
 \includegraphics[width=0.7\textwidth]{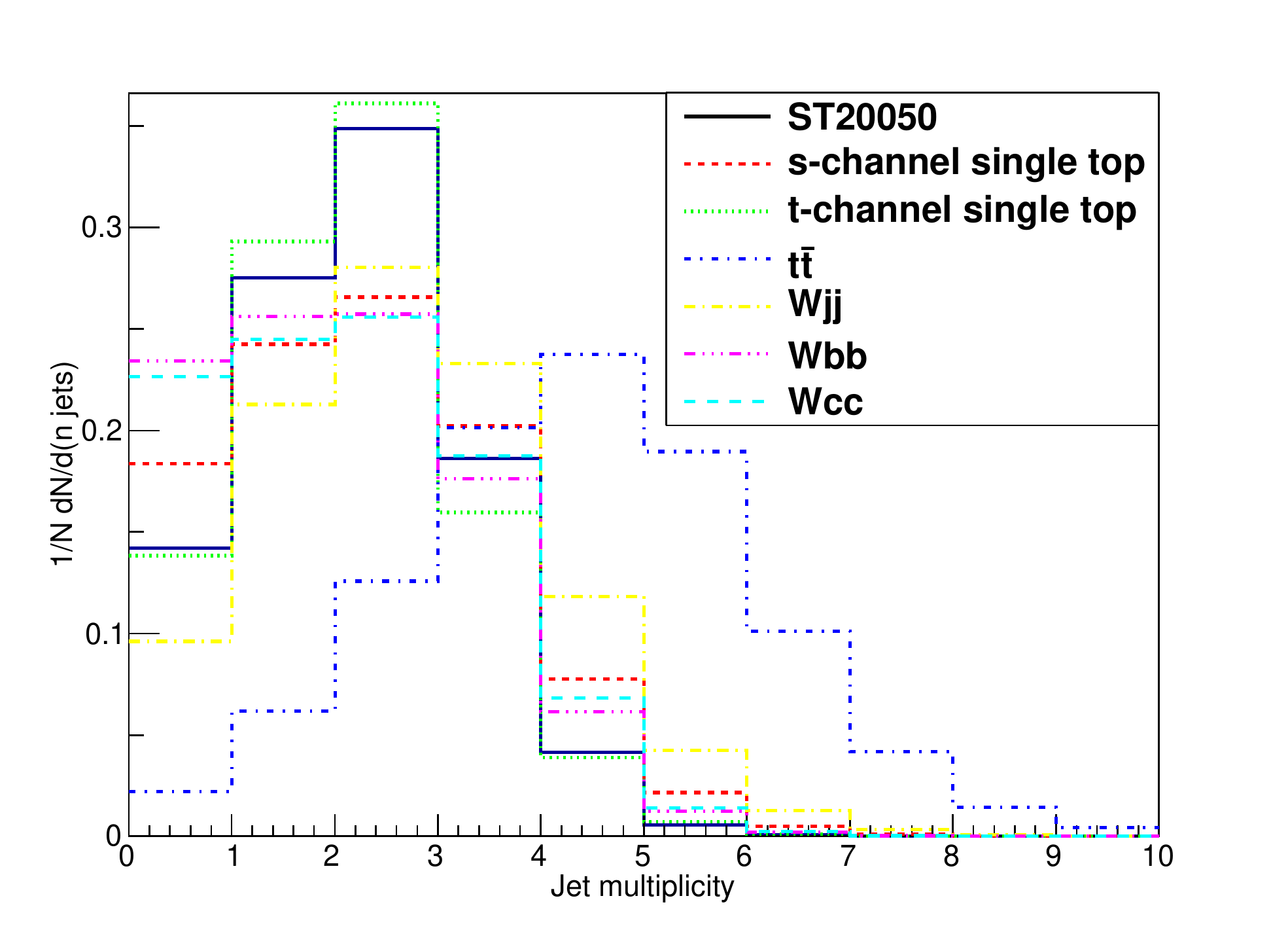}
 \caption{The jet multiplicity distribution in signal and background events. \label{jmul}}
 \end{figure}

\begin{figure}
 \centering
 \includegraphics[width=0.7\textwidth]{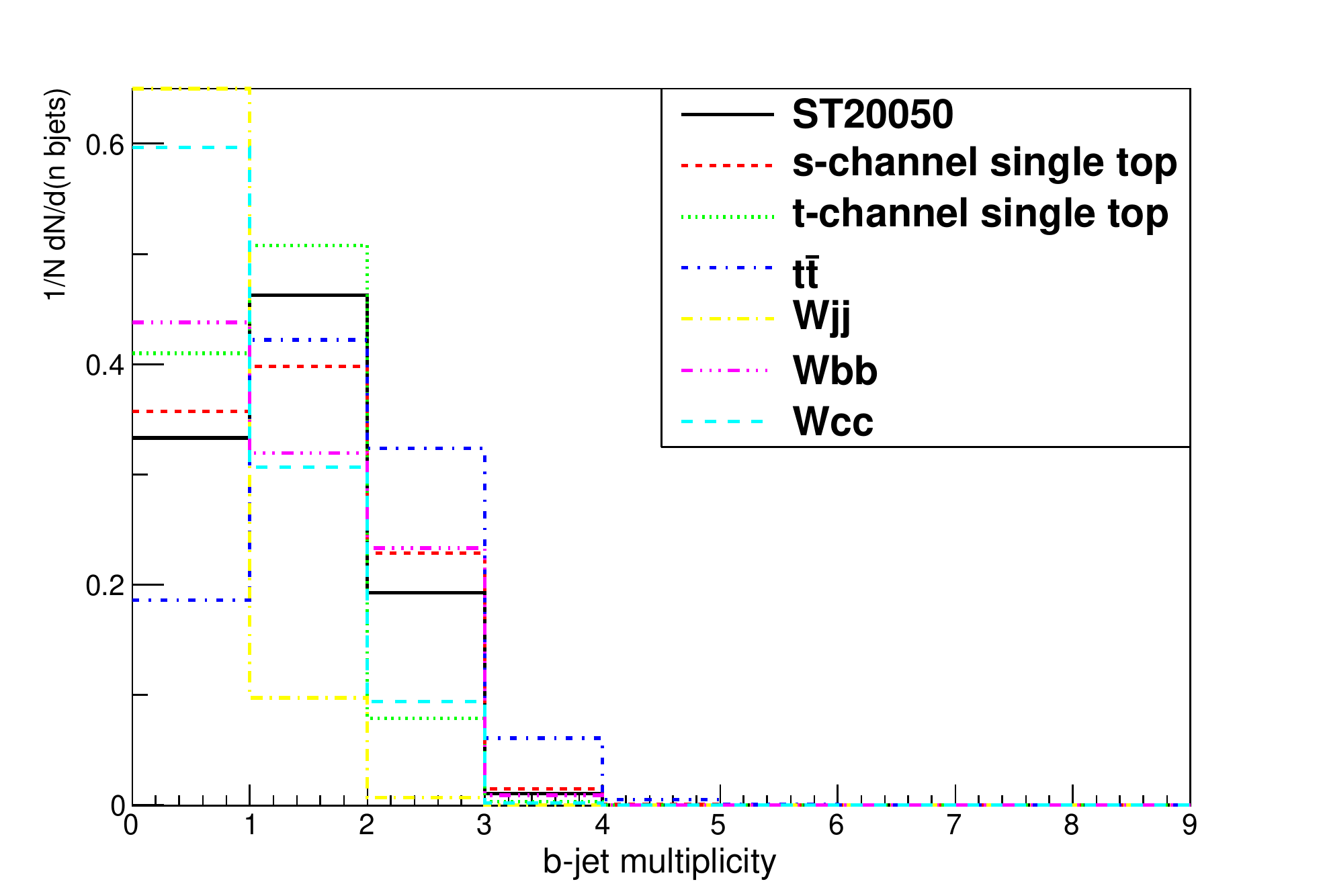}
 \caption{The $b$-jet multiplicity distribution in signal and background events. \label{bjmul}}
 \end{figure}

\begin{table}[h]
\begin{center}
\begin{tabular}{lc}
\hline
Leptons:& $20<p_T^{\textnormal{lepton}}<80$ GeV, $|\eta|<1.5$ \\
\hline
Jets:& $E_T^{\textnormal{jet}}>20$ GeV, $|\eta|<1.5$ \\
\hline
$b$-jets:& $b$-tagging + at least one $b$-jet with $E_T>50$ GeV\\
\hline
$W$ mass window:& $60<\ell\nu$ invariant mass$<90$ GeV\\
\hline
Top quark mass window:& $150<\ell\nu b$ invariant mass$<190$ GeV\\
\hline
$E_T^{\textnormal{miss}}$:& $E_T^{\textnormal{miss}}<70$ GeV\\
\hline  
\end{tabular}
\end{center}
\caption{Selection cuts applied on signal and background events. \label{cuts}}
\end{table}
The kinematic cuts applied on jets are used for $b$-jets with no changes. However, $b$-jets are selected if they match a $b$ or $c$ quark in the event with selection probabilities mentioned before. These requirements lead to the following jet and $b$-jet multiplicity plots shown in Figs. \ref{jmul} and \ref{bjmul}. As mentioned before, only the two jet and two $b$-jet bin is used for the analysis.\\

\subsection{Selection Efficiencies}
\begin{figure}
 \centering
 \includegraphics[width=0.7\textwidth]{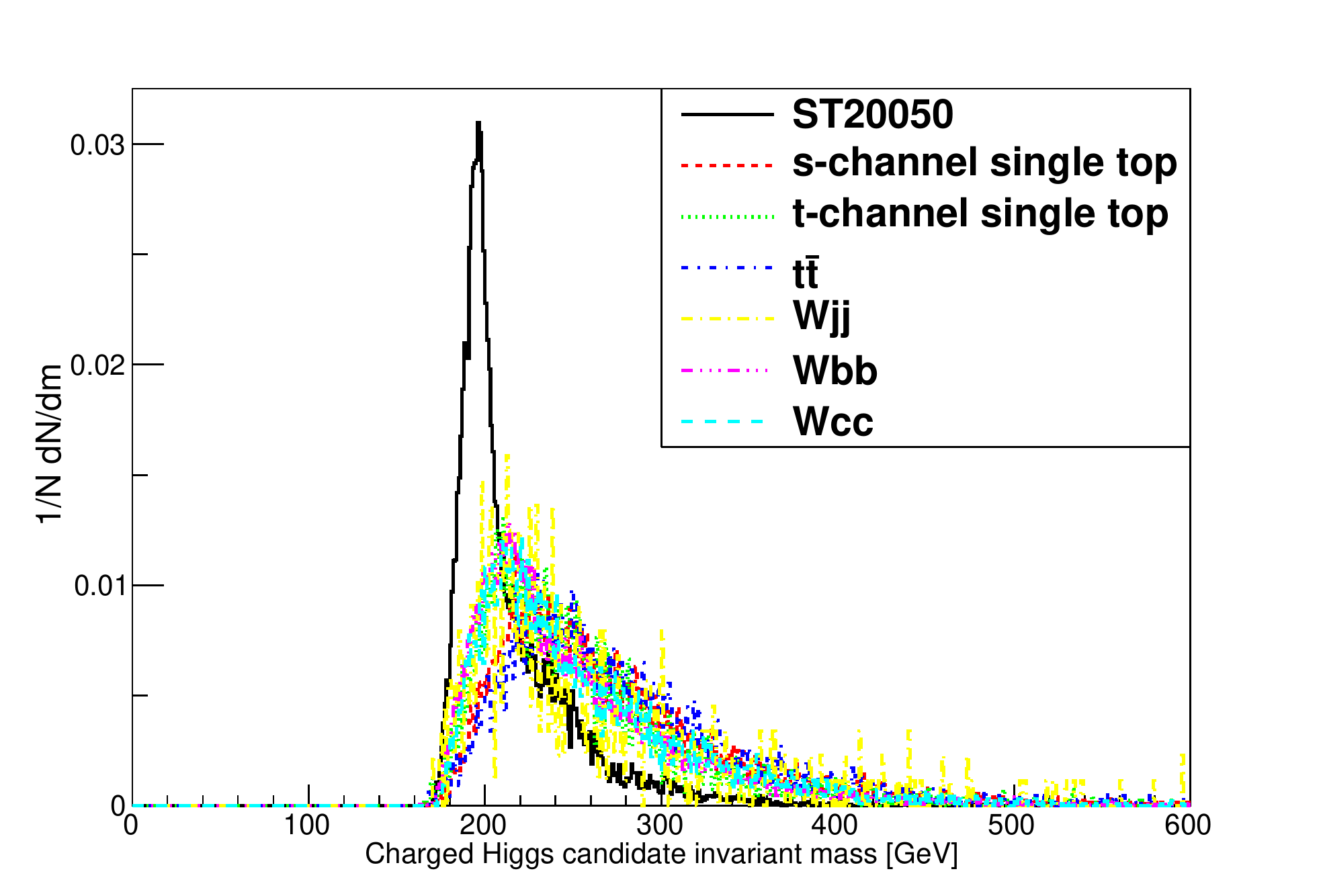}
 \caption{The charged Higgs invariant mass distribution calculated through the $\ell \nu b \bar{b}$ combination. \label{chim}}
 \end{figure}

\begin{figure}
 \centering
 \includegraphics[width=0.7\textwidth]{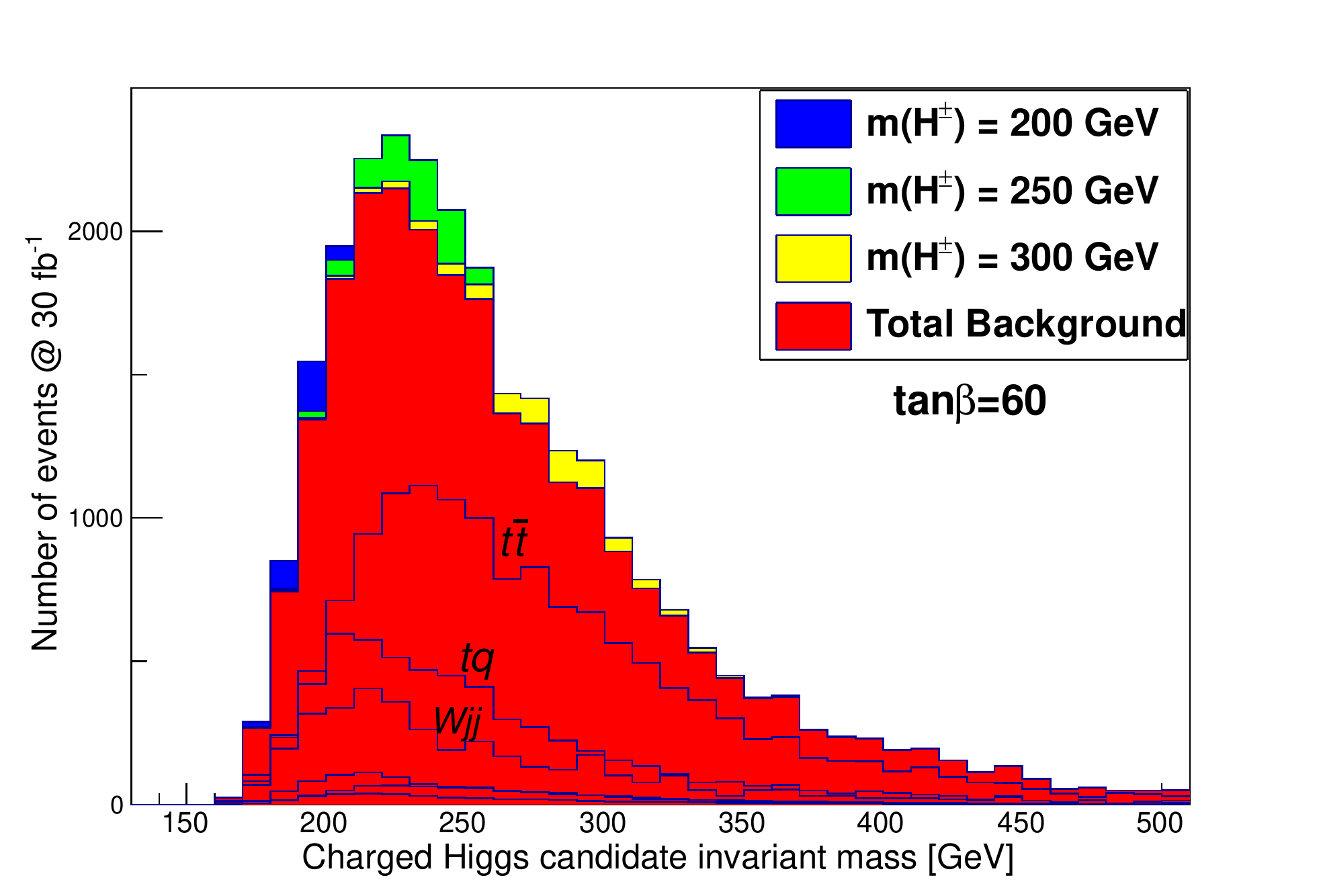}
 \caption{The charged Higgs signal on top of the total background (shown in red). Different charged Higgs mass hypotheses have been shown independently. The main background samples have been labeled. \tanb= 60. \label{chall}}
 \end{figure}

At this stage, signal and background samples pass through the selection cuts and for each cut, a relative efficiency is calculated with respect to the previous cut. Results are quoted in Tab. \ref{signaleff}. As is observed, a charged Higgs with $m_{H^{\pm}}\simeq 180$ GeV is hard to observe due to the very limited phase space available to the top and bottom quarks (resulting in soft kinematic features of the final state particles) and also the low cross section. Selection efficiencies of background events are quoted in Tab. \ref{backgroundeff}. Although the number of background events is large, further reduction is achievable if the charged Higgs candidate invariant mass is reconstructed and a mass window cut is applied on that. The charged Higgs can be reconstructed through $\ell\nu b\bar{b}$ invariant mass reconstruction. Figure \ref{chim} shows the invariant mass of the four particle combination, $\ell\nu b\bar{b}$, which is taken as the charged Higgs candidate. As is seen, the charged Higgs is well reconstructed in signal events, however, there are fake entries related to the background events. Each distribution should of course be normalized to the real number of events at $30fb^{-1}$ including selection efficiencies. The result of such normalization is shown in Fig. \ref{chall} for different charged Higgs masses. A value of \tanb= 60 has been used for illustration. The distribution of $m_{H^{\pm}}=180$ GeV is negligible and thus was not included in the plot. The cut on this distribution depends on the hypothesis on the charged Higgs mass used in the simulation. In other words, the analysis is a mass dependent analysis. The position of the mass window is optimized to obtain the highest possible signal significance. Table \ref{masswindow} shows the mass window coordinates as well as the efficiency of this cut for signal and total background. Also shown are the signal and background events surviving the mass window cut, signal to background ratio, $S/B$, and finally the signal significance, $S/\sqrt{B}$. The $S/B$ reaches roughly 10$\%$ as the best value. 
\begin{table}[h]
\centering
\rotatebox{270}{
\begin{tabular}{|c|c|c|c|c|}
\hline
\multirow{2}{*}{Selection cut} & Signal &Signal &Signal &Signal \\
& $m_{H^{\pm}}=180$ GeV &$m_{H^{\pm}}=200$ GeV &$m_{H^{\pm}}=250$ GeV &$m_{H^{\pm}}=300$ GeV \\
\hline
$\sigma\times$BR [pb] & 0.76 & 1.4 & 1.3 & 0.83 \\  
\hline
One lepton & 15$\%$ & 54$\%$ & 52$\%$& 50$\%$ \\
\hline
Two jets & 41$\%$ & 39$\%$ & 33$\%$& 32$\%$\\
\hline
Two $b$-jets & 2$\%$ & 10$\%$ &23$\%$ &23$\%$\\
\hline
$W$ mass window & 89$\%$ & 94$\%$ & 94$\%$ & 94$\%$\\
\hline
Top quark mass window & 59$\%$ & 77$\%$ & 76$\%$ & 73$\%$ \\
\hline
$E_T^{\textnormal{miss}}$ & 100$\%$ & 98$\%$ & 98$\%$ & 96$\%$ \\
\hline  
Total eff. & 0.06$\%$ & 1.5$\%$ & 2.8$\%$ & 2.4$\%$ \\
\hline
Expected events at $30~fb^{-1}$ & 14 & 630 & 1092 & 598 \\
\hline
\end{tabular}
}
\caption{Signal selection efficiencies with different charged Higgs masses. \tanb= 60 \label{signaleff}}
\end{table}

\begin{table}[h]
\centering
\rotatebox{270}{
\begin{tabular}{|c|c|c|c|c|c|c|}
\hline
\multirow{2}{*}{Selection cut} & \multirow{2}{*}{$t\bar{t}$} & SM single top & SM single top & \multirow{2}{*}{$Wjj$} & \multirow{2}{*}{$Wb\bar{b}$} & \multirow{2}{*}{$Wc\bar{c}$} \\
&  &$s$-channel &$t$-channel & & &  \\
\hline
$\sigma\times$BR [pb] & 227 & 2.1 & 50 & 3340 & 7 & 6.6 \\ 
\hline 
One lepton & 46$\%$ & 39$\%$ & 48$\%$& 30$\%$ & 28$\%$ & 27$\%$\\
\hline
Two jets & 11$\%$ & 31$\%$ & 42$\%$& 32$\%$ & 32$\%$ & 32$\%$\\
\hline
Two $b$-jets & 7$\%$ & 15$\%$ & 3$\%$ & 0.1$\%$ & 11$\%$ & 4$\%$\\
\hline
$W$ mass window & 93$\%$ & 94$\%$ & 95$\%$ & 60$\%$ & 94$\%$ & 95$\%$ \\
\hline
Top quark mass window & 63$\%$ & 71$\%$ & 69$\%$ & 45$\%$ & 52$\%$ & 52$\%$\\
\hline
$E_T^{\textnormal{miss}}$ & 90$\%$ & 94$\%$ & 93$\%$ & 93$\%$  & 92$\%$ & 93$\%$ \\
\hline  
Total eff. & 0.2$\%$ & 1.2$\%$ & 0.4$\%$ & 0.004$\%$ & 0.4$\%$ & 0.2$\%$ \\
\hline
Expected events at $30~fb^{-1}$ & 13620 & 756 & 6000 & 4008 & 840 & 396 \\
\hline
\end{tabular}
}
\caption{Background events selection efficiencies.  \label{backgroundeff}}
\end{table}


\begin{table}[h]
\centering
\begin{tabular}{|c|c|c|c|c|c|c|}
\hline
\multirow{2}{*}{Sample} & \multicolumn{2}{|c|}{Mass window} & \multirow{2}{*}{Eff.} &\multirow{2}{*}{Events}& \multirow{2}{*}{S/B} & \multirow{2}{*}{S/$\sqrt{B}$} \\
& Lower limit & Upper limit & & & &  \\
\hline
Signal, $m_{H^{\pm}}=180$ GeV &  180 & 300 & 97$\%$ & 14 & \multirow{2}{*}{0.0008} & \multirow{2}{*}{0.1} \\
Total Background & 180 & 300 & 76$\%$ & 19014 &  & \\ 
\hline
Signal, $m_{H^{\pm}}=200$ GeV &  180 & 210 & 61$\%$ & 389 & \multirow{2}{*}{0.09} & \multirow{2}{*}{6} \\
Total Background & 180 & 210 & 17$\%$ & 4187 &  & \\ 
\hline
Signal, $m_{H^{\pm}}=250$ GeV &  200 & 260 & 80$\%$ & 870 & \multirow{2}{*}{0.07} & \multirow{2}{*}{7.6} \\
Total Background & 200 & 260 & 52$\%$ & 13078 &  & \\ 
\hline
Signal, $m_{H^{\pm}}=300$ GeV &  260 & 320 & 71$\%$ & 433 & \multirow{2}{*}{0.05} & \multirow{2}{*}{4.7} \\
Total Background & 260 & 320 & 33$\%$ & 8325 & & \\ 
\hline
\end{tabular}
\caption{Charged Higgs candidate mass window cuts and final results on the signal to background ratio and the signal significance. The quoted efficiencies are the mass window cut efficiencies for signal and background samples. For $m_{H^{\pm}}=180$ GeV, a large window has been used due to the low signal statistics. \tanb= 60. \label{masswindow}}
\end{table}

\section{Results}
\begin{figure}
 \centering
 \includegraphics[width=0.7\textwidth]{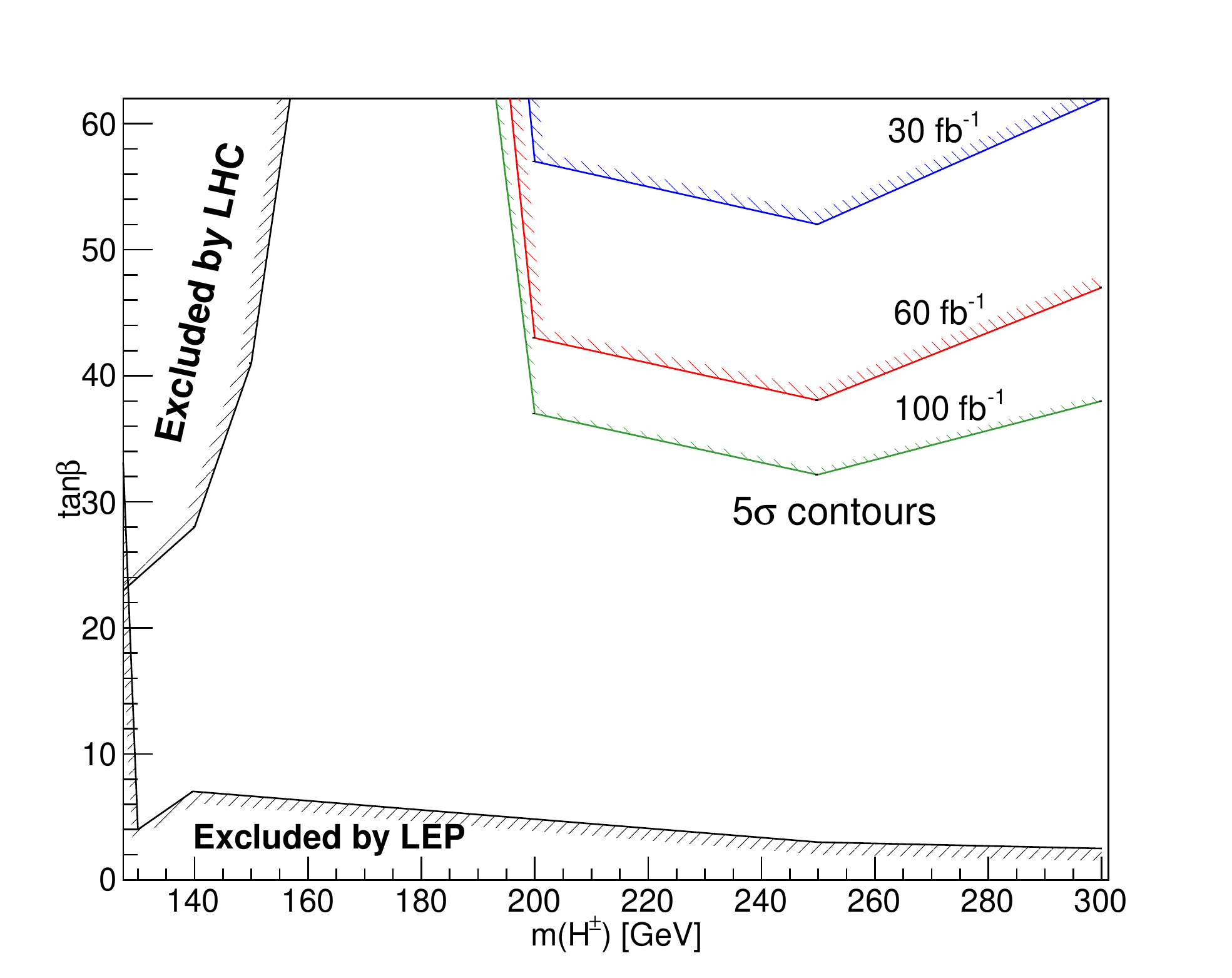}
 \caption{The $5\sigma$ discovery contour using the $s$-channel single top as a source of heavy charged Higgs boson. Results are shown for different integrated luminosities of LHC operating at $\sqrt{s}=14$ TeV. The already excluded areas of LEP and LHC are also included. \label{5s}}
 \end{figure}

\begin{figure}
 \centering
 \includegraphics[width=0.7\textwidth]{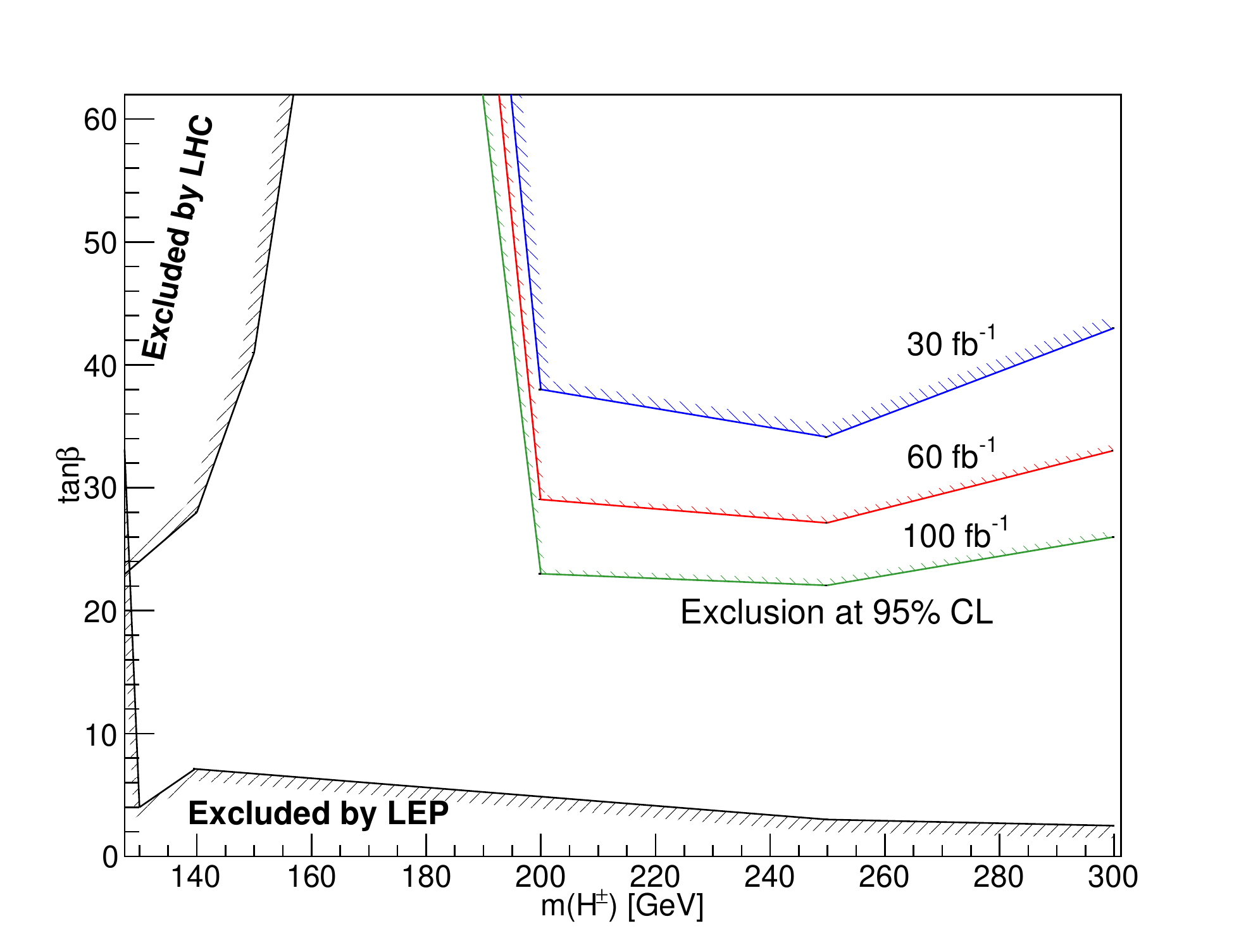}
 \caption{The $95\%$ C.L. exclusion contour using the $s$-channel single top as a source of heavy charged Higgs boson. Different integrated luminosities of LHC, operating at $\sqrt{s}=14$ TeV, as well as the already excluded areas of LEP and LHC are included.\label{95CL}}
 \end{figure}

The number of signal events mentioned in the previous section was calculated for a specific \tanb value. i.e., \tanb= 60. In order to obtain a contour of $5\sigma$ discovery or exclusion at 95$\%$ C.L., other \tanb values are studied. The TLimit class implemented in ROOT \cite{root} is used for both contours. The results are shown in Figs. \ref{5s} and \ref{95CL}. The current excluded areas by LEP and LHC are also shown. Different integrated luminosities are included for comparison. These results show that a wide range of parameter space is available for discovery or exclusion at integrated luminosities which will be available soon at LHC. Considering current results of LHC, this analysis shows that using single top events helps a lot in the search for this particle in the heavy mass area.  

\section{Conclusions}
The $s$-channel single top production process was used as a source of heavy charged Higgs boson at LHC. An analysis of signal and background selection was designed by simulating LHC events at $\sqrt{s}=14$ TeV. The charged Higgs reconstruction was performed as a tool for increasing the signal to background ratio and the signal significance. Comparing different integrated luminosities, it is concluded that a charged Higgs boson can well be observed or excluded in a wide range of ($m_{H^{\pm}},$\tanb) parameter space. In order to demonstrate the results, $5\sigma$ discovery and $95\%$ CL exclusion contours are presented for different integrated luminosities. Nevertheless this process requires large \tanb > 30 and it should also be noted that the signal significance was computed neglecting systematic and theoretical uncertainties on the backgrounds. 

 
\pagebreak

\end{document}